\documentclass[fleqn,twoside]{article}
\usepackage{espcrc2}
\usepackage{epsfig,graphics}

\usepackage{graphicx}
\usepackage[figuresright]{rotating}

\newcommand{\be}{\begin{equation}}
\newcommand{\ee}{\end{equation}}

\newcommand{\AmS}{{\protect\the\textfont2
  A\kern-.1667em\lower.5ex\hbox{M}\kern-.125emS}}

\title{Glueballs, strings and topology in SU(N) gauge theory} 

\author{M. Teper\address{Theoretical Physics, University of
        Oxford, \\
        1 Keble Road, Oxford OX1 3NP, United Kingdom}
        \thanks{Invited talk given at the Workshop on Lattice Hadron Physics 
        (LHP2001), July 9-18, 2001 in Cairns, Australia.}
}

\begin{document}

\begin{abstract}
I show how one can use lattice methods to calculate various continuum
properties of SU($N$) gauge theories; in part to explore old ideas that
$N=3$ might be close to $N=\infty$. I describe  calculations of 
the low-lying `glueball' mass spectrum, of the string tensions of
$k$-strings and of topological fluctuations for  $2 \leq N \leq 5$.
We find that mass ratios appear to show a rapid approach to the 
large--$N$ limit, and, indeed, can be described all the way down to 
SU(2) using just a leading $O(1/N^2)$ correction. We confirm that the 
smooth large--$N$ limit we find is confining and is obtained by
keeping a constant 't Hooft coupling. We find that the ratio 
of the $k=2$ string tension to the $k = 1$ fundamental string 
tension is much less than the naive (unbound) value of 2 and is
considerably greater than the naive bag model prediction; in 
fact it is consistent, within quite small errors, with either
the M(-theory)QCD-inspired conjecture that 
$\sigma_{k} \propto \sin(\pi k/N)$ or with `Casimir scaling'. 
Finally I describe calculations of the topological charge of the 
gauge fields.
We observe that, as expected, the density of small-size instantons 
vanishes rapidly as $N$ increases, while the topological susceptibility 
appears to have a non-zero $N = \infty$ limit.

\vspace{1pc}
\end{abstract}

% typeset front matter (including abstract)
\maketitle

\section{INTRODUCTION}
\label{sec_introduction}

How SU($N$) gauge theories approach their $N=\infty$
limit and what that limit is, are interesting questions
\cite{largeN},
in themselves,
whose answers would, in addition, represent a significant step 
towards addressing the same question in the context of QCD.
Accurate lattice calculations in 2+1 dimensions 
\cite{mt98}
show that in that case the approach is remarkably precocious 
in that even $N=2$ is close to $N=\infty$. Such calculations 
have to be very accurate because for each 
value of $N$ one has to perform a continuum extrapolation
of various mass ratios and then these are compared and 
extrapolated to $N=\infty$. Earlier D=3+1 calculations 
\cite{mtoldN,winoh}
were much too rough for this purpose even if their message
was optimistic. Recently, however, 
this situation has been rapidly improving and here I 
want to describe some of the things that one has 
already learned.

Analyses of Feynman diagrams to all orders imply
\cite{largeN},
that the SU($N$) theory will have a smooth limit if
we keep fixed the 't Hooft coupling
\begin{equation}
\lambda \equiv g^2 N.
\end{equation}
We would like a non-perturbative confirmation of this
expectation. Again, we would like a confirmation of the
assumption that the large-$N$ theory remains confining,
since much of the phenomenological argument that QCD might 
be close to $N=\infty$ relies on this being so
\cite{largeN}.
Since quarks are in the fundamental representation, the
leading correction to $N=\infty$ is expected to be
$O(1/N)$. In the pure gauge theory it is expected to be
$O(1/N^2)$. (Again an expectation that one would wish to
test non-perturbatively.) Thus if QCD is close to $N=\infty$ 
one would expect the SU(3) gauge theory to be close to
SU($\infty$) as well. We can test this by calculating
dimensionless ratios of physical masses as a function
of $N$. There are also arguments and speculations
(e.g. \cite{witten})
that topological fluctuations are very different at large
$N$ and at small $N$; this too we would like to  
test in lattice simulations.

The calculations 
\cite{blmtglue,blmtstring}
on glueballs, topology and $k$-strings
which I will describe have been performed in collaboration 
with Biagio Lucini. The strategy is simple:
we calculate the relevant (continuum) properties of
SU(2), SU(3), SU(4) and SU(5) theories and directly
compare them. Since $N=5$ is large in the sense
that the leading correction $\propto 1/N^2 = 1/25 \ll 1$,
it is not surprising that such a calculation turns out to 
suffice for our purposes. These
calculations were intended as exploratory. A much
larger calculation is now underway and this will
provide much more information on the mass spectrum
and its dependence on the number of colours.

\section{FROM LATTICE TO PHYSICS}
\label{sec_fromlatticetophysics}

Our lattice calculations are entirely standard
\cite{blmtglue,blmtstring}.
The Euclidean space-time lattice is hypercubic with periodic 
boundary conditions.The degrees of freedom are SU($N$) 
matrices residing on the links of the lattice.
In the partition function the fields are weighted with
$\exp\{S\}$ where $S$ is the standard plaquette action
\be
S = - \beta \sum_p \biggl (1 - {1\over N} ReTr \, U_p \biggr ),
\label{eqn_action}
\ee
in which $U_p$ is the ordered product of the matrices on
the boundary of the plaquette $p$. For smooth fields this
action reduces to the usual continuum action with
$\beta = 2N/g^2$. For rough fields on a lattice of spacing
$a$ we can define a running lattice coupling $g_L(a)$ which 
reduces in the continuum limit to 
a coupling $g(a)$ in our favourite scheme:
\be
\beta \equiv {{2N}\over{g_L^2(a)}}
\stackrel{a\to0}{\longrightarrow} {{2N}\over{g^2(a)}}
\label{eqn_beta}
\ee
Thus by varying the inverse lattice coupling $\beta$ we 
can vary the lattice spacing $a$.  

To calculate a mass we construct some operator $\phi(t)$
with the quantum numbers of the state (typically this will
be a linear combination of the products of link matrices 
around some closed loops) and then use the
standard decomposition of the Euclidean correlator in
terms of energy eigenstates
\be
C(t) = \langle \phi^{\dagger}(t) \phi(0) \rangle
= 
\sum_n | \langle \Omega | \phi | n \rangle |^2
\exp \{- E_n t \}
\label{eqn_corrln}
\ee
where $|n\rangle$ are the energy eigenstates, with $E_n$ the 
corresponding energies, and $|\Omega\rangle$ is the vacuum state. 
We evaluate the corresponding Feynman Path Integrals using
standard Monte Carlo techniques. On the lattice $t=na$ and
so we obtain the energies from eqn(\ref{eqn_corrln}) as
$aE_n$ i.e. in units of the lattice spacing. In practice
one needs to use several carefully chosen operators, and
a variational calculation 
\cite{blmtglue,blmtstring}.
Since our volume is finite we must make sure that finite volume
corrections are negligible. For the calculations I describe
here such checks have been made. 

Having calculated some masses $am_i$ at a fixed value of $a$
(i.e. at a fixed value of $\beta$) we can remove lattice units
by taking ratios: $am_i/am_j = m_i/m_j$. This ratio
differs from the desired continuum value by lattice corrections.
For our action the functional form of the leading correction is
known to be $O(a^2)$. Thus for small enough $a$ we can 
extrapolate our calculated mass values
\be
{{m_i(a)} \over {m_j(a)}} =
{{m_i(0)} \over {m_j(0)}} 
+ c a^2 m^2_k(a)
\label{eqn_cont}
\ee
where using different $m_k$ will make differences at $O(a^4)$
(as does the $a$-dependence of $m_k(a)$). We shall later show some
explicit examples of such extrapolations.

Having calculated such  mass ratios in various continuum SU($N$)
gauge theories, we can attempt to extrapolate to $N=\infty$ 
using the fact that the leading correction is expected to
be $O(1/N^2)$:
\be
\left.  {{m_i} \over {m_j}} \right |_{N} =
\left.  {{m_i} \over {m_j}} \right |_{\infty} 
+ {c_{ij} \over N^2}.
\label{eqn_largeN}
\ee
If such fits are good for $N\geq N_0$ then we may regard
SU($N_0$) as being close to SU($\infty$).

\section{GLUEBALLS}
\label{sec_glueballs}

\begin	{figure}
\begin	{flushleft}
\epsfig{figure=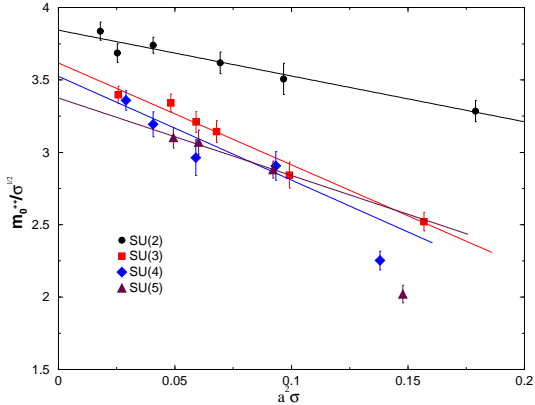, angle=270, width=7cm} 
\end	{flushleft}
\caption{The mass of the lightest scalar glueball, $m_{0^{++}}$, 
expressed in units of the string tension, $\sigma$, is plotted 
against the latter in lattice units. The $a\to 0$ continuum 
extrapolation, using a leading lattice correction, is shown.}
\label{fig_scalar}
\end 	{figure}

In this exploratory calculation we focus on what are
expected (from previous calculations in SU(2) and SU(3)
\cite{MTrev98,MPanisotropic})
to be the lightest states: the lightest 
$J^{PC} = 0^{++}$ and $2^{++}$ glueballs. We also
calculate the mass of the first excited scalar glueball,
which we shall refer to as the $0^{++\star}$.
In addition to these glueball masses we calculate
the string tension, $a^2\sigma$, of the flux tube
between static sources in the fundamental representation
(see the next Section.). This last is our most accurately
calculated quantity and so we use it to form the
dimensionless mass ratios $m/\surd\sigma$ which
we extrapolate to the continuum limit using
eqn(\ref{eqn_cont}). In Fig.\ref{fig_scalar} I show you how 
this works for the lightest scalar glueball for $N=2,3,4,5$.

\begin	{figure}
\begin	{flushleft}
\epsfig{figure=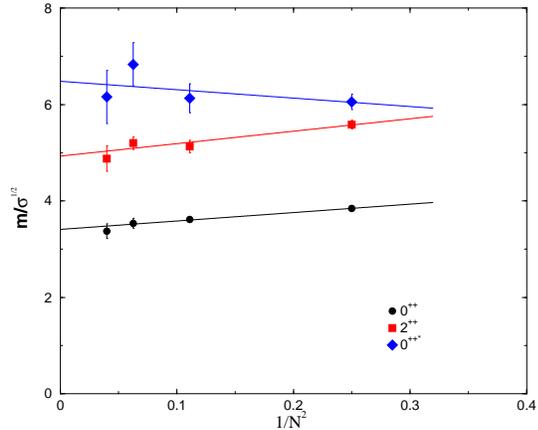, angle=270, width=7cm} 
\end	{flushleft}
\caption{Continuum scalar, tensor and excited scalar
masses expressed in units of the string tension and plotted
against $1/N^2$. Linear extrapolations to $N=\infty$
are shown in each case.} 
\label{fig_glueN}
\end 	{figure}

While the mass of the scalar glueball is the most accurately
calculated, because it is the lightest, it is also the one
with (by far) the largest lattice corrections. Despite this
it is already clear from Fig.\ref{fig_scalar} that for small $a$
there is very little $N$-dependence for $N\geq 3$. 
To be quantitative we perform continuum extrapolations
using eqn(\ref{eqn_cont}), as shown in  Fig.\ref{fig_scalar}.
The results are shown in Fig.\ref{fig_glueN}
where we also show the extrapolations to $N=\infty$
using  eqn(\ref{eqn_largeN}). 

We see from Fig.\ref{fig_glueN} that, as far as the lightest
glueballs are concerned, all SU($N$) theories can be described
by a modest leading $O(1/N^2)$ correction to the  SU($\infty$)
limit. 

We also see from these plots that the ratio of the glueball
masses to the (square root of) the string tension has
a finite non-zero limit as $N \to \infty$. This tells us
that the confining string tension remains finite and
non-zero as  $N \to \infty$. (Caution: numerical calculations
only test confinement to some finite distance -- which in
the case of our higher-$N$ calculations is not yet very large.)

\section{'T HOOFT COUPLING}
\label{sec_coupling}

The analysis of diagrams 
\cite{largeN}
suggests that the way to achieve a smooth limit as $N\to\infty$ 
is by keeping the 't Hooft coupling, $\lambda \equiv g^2 N$, constant.
Since the coupling runs, we should say that 
what we keep fixed is the running
't Hooft coupling, as defined on some scale $l$ that is fixed in units 
of some quantity that partakes of the smooth large-$N$ limit,
such as the string tension. To do this we use eqn(\ref{eqn_beta})
which tells is that a suitable defintion of a running 't Hooft
coupling is 
\be
\lambda_I(a) =  g_I^2(a) N 
= {{2N^2}\over{\beta \langle ReTr \, U_p /N \rangle}}
\label{eqn_lambdaI}
\ee
The extra factor involving the plaquette is a mean-field (or tadpole)
improved version of $\beta$ and the naive $\lambda(a)$ we would derive 
from it
\cite{gimp}.
It is necessary
\cite{gimp}
because the naive lattice coupling is known to be very poor in the
sense of having very large higher order corrections.

\begin	{figure}
\begin	{flushleft}
\epsfig{figure=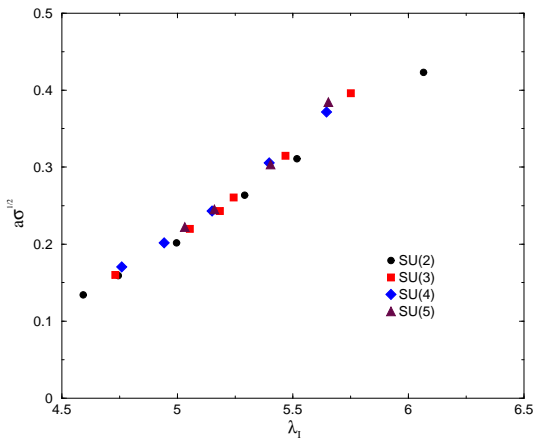, angle=270, width=7cm} 
\end	{flushleft}
\caption{The square root of the string tension in lattice units,
$a\surd\sigma$, plotted against the 't Hooft coupling, 
$\lambda_I \equiv g^2_I N$.}
\label{fig_betaI}
\end 	{figure}

We extract $\lambda_I(a)$ for each of our various lattice calculations 
at various $N$ and $a$ . Our diagrammatic expectation is that if 
we plot  $a\surd\sigma$ against $\lambda_I(a)$ then, for large enough 
$N$, the calculated values should fall on a universal curve.
In Fig.\ref{fig_betaI} we plot all our calculated values of 
$a\surd\sigma$,  against the value of  $\lambda_I(a)$.
We observe that, within corrections which are small except at
the largest value of $a$, we do indeed see a universal curve,
and so the diagrammatic expectation is supported by our 
non-perturbative calculations  
\cite{blmtglue}.
\section{STRINGS}
\label{sec_strings}

In a confining SU($N$) gauge theory, the potential between two
static charges in the fundamental representation is expected
to grow linearly at large separations as
$V(r) = \sigma r$ where $\sigma$ is the fundamental string tension.
If we have charges in other representations then, in SU(3), they 
can be screened to either the fundamental or the trivial representation
by gluons from the vacuum. So the  fundamental string is the
only stable one (assuming it to have the lowest tension).
In SU($N\geq 4$) there are other stable strings. It is convenient
to label a charge by the integer $k$ if its wave function transforms
by a factor of $z^k$ under a global gauge transformation $z$ 
belonging to the centre $Z_N$ of the group. Because gluons
transform trivially under the centre, they cannot screen
a charge of $k$ into a charge of $k^\prime$ if
$z^k \neq z^{k^\prime}$. Thus there should be a different stable string
for each such $k$, with string tension $\sigma_k$. (Conjugate strings 
will have the same tensions.) Of course, it might be that we
just have $k$ fundamental strings joining such sources, in
which case one will find $\sigma_k = k \sigma$ and we have
no new string. As we shall see, this is not the case: we find new
tightly bound $k$-strings. The first $k=2$ string appears in SU(4)
and the first $k=3$ one in SU(6). The values of  $\sigma_k$ are
interesting because they carry information about confinement.
Also because there are specific conjectures about their values
from M(-theory)QCD
\cite{MQCD},
from arguments about Casimir scaling
\cite{casimir},
and from certain models, such as the bag model
\cite{bag}.
Moreover such strings may have striking implications
\cite{blmtglue,blmtstring}
for the glueball mass spectrum as a function of $N$.

\begin	{figure}[p]
\begin	{flushleft}
\leavevmode
\epsfig{figure=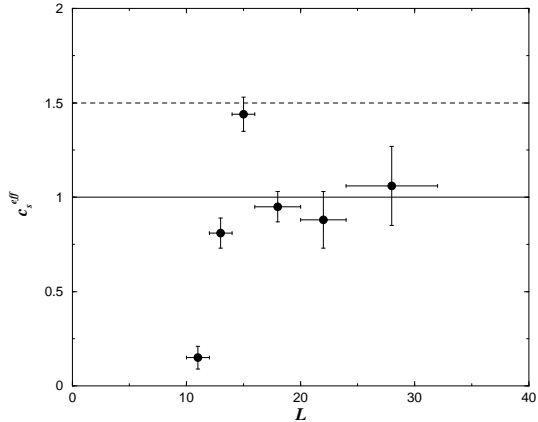, angle=270, width=7cm} 
\end	{flushleft}
\caption{The D=3+1 effective string correction coefficient estimated 
from the masses of flux loops of different lengths (indicated
by the span of the horizontal error bar).
The solid line is what one expects for
a simple bosonic string. For comparison the dashed line indicates the
value for the Neveu-Schwartz string.}
\label{fig_d4cspair}
\end 	{figure}

The simplest way to calculate $\sigma_k$ is by calculating 
the mass of a $k$-string that winds around a spatial torus.
In a confining theory it cannot break and will have a
length $l=aL$ (on an $L^3$ spatial lattice). For long enough
strings, the mass of such a loop is given by
\be
m_k(l) \stackrel{l\to\infty}{\simeq} 
\sigma_{k} l - 
{{\pi(D-2)}\over{6}}{c_s\over l}.
\label{eqn_poly}
\ee
where the $O(1/l)$ correction is the Casimir energy of a periodic
string and $c_s$ is proportional to the central 
charge. This correction is universal 
\cite{Luscherstring}
since it depends only upon the massless modes in the effective 
string theory and does not depend upon the detailed
and complicated dynamics of the flux tube on scales comparable 
to its width. The central charge is given 
\cite{Polchinski}
by the number of massless bosonic and fermionic modes 
that propagate along the string. In practice it is
usually assumed that $c_s=1$, corresponding to the simplest 
possible (Nambu-Goto) bosonic string theory. However, these
modes are not related to the fundamental degrees of freedom
of our SU($N$) gauge theory in any transparent way and the presence
of fermionic modes is certainly not excluded. Direct numerical
evidence is hard to get since we are interested in small corrections
to long, massive strings. In Fig.\ref{fig_d4cspair} we show the
effective value of $c_s$ that one obtains
\cite{blmtstring}
by fitting the masses
of two strings of different lengths (as indicated) to
eqn(\ref{eqn_poly}) in SU(2) and for a lattice spacing
$a \simeq 0.16/\surd\sigma \simeq 0.07 fm$. This provides some 
evidence for the simple bosonic string correction for lengths 
$l\geq 1.2 fm$. 

\begin	{figure}[p]
\begin	{flushleft}
\leavevmode
\epsfig{figure=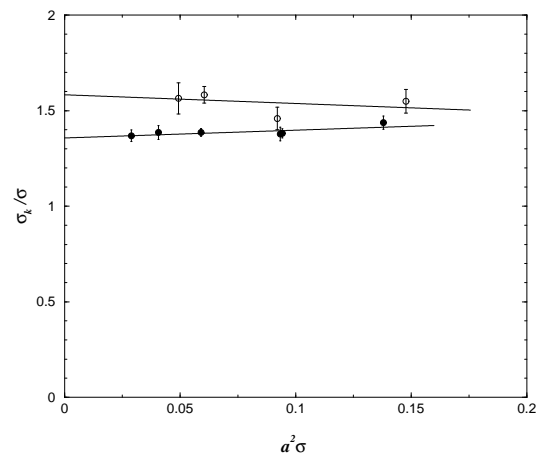, angle=270, width=7cm} 
\end	{flushleft}
\caption{The ratio of $k=2$ and $k=1$ string tensions in 
our D=3+1 SU(4) ($\bullet$) and SU(5) ($\circ$) lattice
calculations plotted as a function of $a^2\sigma$.
Extrapolations to the continuum limit, using a leading 
$O(a^2)$ correction, are displayed.}
\label{fig_d4sig2}
\end 	{figure}

Assuming, then, eqn(\ref{eqn_poly}) with $c_s = 1$ we obtain
from our calculated loop masses
the string tensions shown in Fig.\ref{fig_d4sig2},
from which we obtain
\be
\lim_{a \to 0}
{{\sigma_{\mathit{k=2}}} \over {\sigma}}
=
\left\{ \begin{array}{ll}
1.357 \pm 0.029  & \ \ \ {\mathrm{SU(4)}} \\
1.583 \pm 0.074  & \ \ \ {\mathrm{SU(5)}}
\end{array}
\right. .
\label{eqn_contd4}
\ee
Clearly the $k=2$ string is tightly bound, and the string
tension is incompatible, for example, with the naive bag
model. What we find is that it falls between the MQCD
and Casimir scaling predictions and, at the two sigma
level, is consistent with both. We remark that a very
recent higher statistics calculation
\cite{Pisa}
favours MQCD over Casimir scaling.

\section{TOPOLOGY}
\label{sec_topology}

Gauge fields in four (suitably compactified) Euclidean dimensions 
possess non-trivial topological properties characterised by an 
integer topological charge $Q$. Topological fluctuations break the 
(approximate) $U_A(1)$ symmetry of QCD, thus leading 
\cite{eta-thooft}
to the non-Goldstone character of the $\eta^\prime$.
One can argue that if $N=3$ is close to $N=\infty$
one can relate 
\cite{WV}
the mass of the $\eta^\prime$ to the topological susceptibility
\be
\chi_t \equiv {{\langle Q^2} \rangle \over V},
\label{eqn_chi}
\ee
of the gauge theory without quarks. ($V$ is the space-time volume.)
This suggests a value $\chi_t \simeq (180 MeV)^4$, which is indeed
close to what one finds in pure gauge theories
\cite{teperPisa}.
All this fits together if we can show that the $N=3$ values of
the pure gauge susceptibility and of $Nm^2_{\eta^\prime}$ are close 
to  their $N=\infty$ values. While we cannot say much about the 
latter, we can and shall address the former.

\begin	{figure}
\begin	{flushleft}
\epsfig{figure=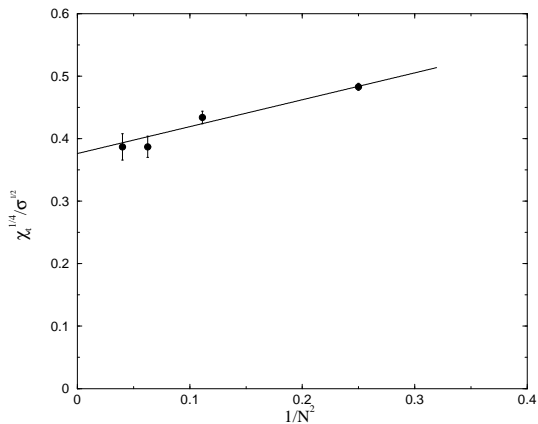, angle=270, width=7cm} 
\end	{flushleft}
\caption{The continuum topological susceptibility in units 
of the string tension plotted against $1/N^2$. A linear
extrapolation to $N=\infty$ is shown.}
\label{fig_khiN}
\end 	{figure}

We calculate $Q$ for our lattice fields by standard cooling methods 
which for this purpose are reliable
\cite{teperPisa}.
These calculations are performed simultaneously with the
glueball calculations discussed earlier and the values of
$\chi^{1/4}_t/\surd\sigma$ are extrapolated to the continuum
limit with an $a^2\sigma$ correction, following
eqn(\ref{eqn_cont}). The resulting continuum values are plotted
against $1/N^2$ in Fig.\ref{fig_khiN}
We observe that, just as for the glueball masses, the $N$-dependence
is well described by a modest $O(1/N^2)$ correction 
and the topological susceptibility, expressed in physical
units, does not differ greatly when we go from $N=3$ to
$N=\infty$.

These conclusions are somewhat weaker here than for the glueballa
because of the rapidly growing errors as $N$ increases.
The reason for this turns out to be interesting and easy to see
\cite{blmtglue}.
As $N$ grows isolated instantons become increasingly unlikely.
This is because of the factor in the instanton density 
\be
D(\rho) d\rho
\propto 
{{d\rho}\over{\rho}}{1\over{\rho^4}}
\Biggl\{
{{b^2}\over{\lambda^2(\rho)}}
e^{-{{8\pi^2}\over{\lambda(\rho)}}}
\Biggr\}^N .
\label{eqn_MTI}
\ee
where $\lambda$ is the 't Hooft coupling that we keep fixed
as $N\to\infty$. In the real vacuum this argument holds
for instantons with $\rho\ll 1$ `fermi' and so these
instantons  are exponentially suppressed as $N$ increases.
Now the Monte Carlo changes $Q$ by an instanton shrinking through
small values of $\rho$ down to $\rho \sim a$ where
it can vanish through the lattice. (Or the reverse process.) 
However eqn(\ref{eqn_MTI}) tells us that the probability
of a very small instanton goes rapidly to zero as $N$
grows. Thus the lattice fields rapidly become constrained 
to lie in given topological sectors and for this quantity
the Monte Carlo rapidly ceases to be ergodic as $N$ grows,
and the stattistical errors on $\langle Q^2 \rangle$
grow rapidly -- as observed.

\begin	{figure}
\begin	{flushleft}
\epsfig{figure=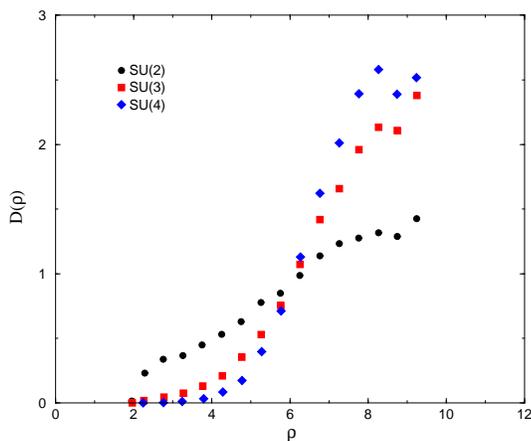, angle=270, width=7cm} 
\end	{flushleft}
\caption{The number density of topological charges plotted 
as a function of the charge radius, $\rho$; obtained
for $N=2,3,4$ from $20^4$ lattices with 
$a\surd\sigma \sim 0.16$.}
\label{fig_drho}
\end 	{figure}

We can attempt to calculate $D(\rho)$ and so see this effect directly.
To do so we examine the topological charge density on the cooled fields 
and associate an instanton to each peak in this density using
the semiclassical formula
\be
Q_{peak} = {6\over{\pi^2\rho^4}}.
\label{eqn_Qpeak}
\ee
This procedure is clearly of ambiguous validity for broad instantons, 
but is unambiguous for the high narrow peaks that correspond to the
smaller instantons of interest to us here. In Fig.\ref{fig_drho}
we display the result of such a calculation, comparing the 
instanton densities in SU(2), SU(3) and SU(4) on lattices
with (almost) equal volumes and lattice spacings when expressed
in units of the string tension. We clearly see a drastic suppression
of small instantons with increasing $N$.
\section{CONCLUSIONS}
\label{sec_conclusions}

We saw above that, as far as the lightest glueballs, string
tensions and topological susceptibility are concerned, all SU($N$) 
theories can be described by a modest leading $O(1/N^2)$ correction 
to the  SU($\infty$) limit. In this sense we can say that
not only is $N=3$ close to $N=\infty$ but so is $N=2$. 

The large amount of interesting physics delivered by these modest 
(workstation) calculations, provides a strong motivation for going 
further. Currently we 
\cite{newglue}
are starting much larger calculations
on anisotropic lattices, which will enable us to obtain accurate
mass estimates for a much larger range of glueball states and
very accurate string tension ratios. Simultaneously we
\cite{newgw}
are using overlap fermions
\cite{neuberger}
to determine the relationship between topology and chiral symmetry
breaking as a function of $N$; and to say something about
instantons and topology at large $N$. A more ambitious project,
which would require the use of one of the larger $\sim 1$ teraflop 
resources that are becoming available to many groups, would 
be to calculate the quenched hadron spectrum at large $N$ -- which
is interesting since it approaches the correct spectrum of SU($N$) QCD 
as we take $N\to\infty$ at fixed non-zero quark mass. And then,
eventually, to dynamical quarks at large $N$ and the $\eta^\prime$ ...

\section*{Acknowledgements}

I am very grateful to the organizers for inviting me to participate
in this interesting (and enjoyable) workshop, and to my fellow 
participants for many useful comments and discussions.

\end{document}